\documentclass[twocolumn]{aastex6}

\usepackage{times,color,natbib,url,amsmath}
\usepackage{graphicx,float,psfrag}
\usepackage{tabularx}
\usepackage{latexsym,amsmath,amssymb}
\usepackage{natbib}
\usepackage{verbatim}
\usepackage{multirow}
\usepackage{color}

\bibpunct{(}{)}{;}{a}{}{,}

\newcommand {\asca} {{\it ASCA}}

\newcommand {\sax} {{\it BeppoSAX}}

\newcommand {\xmm} {\textsl{XMM-Newton}}
\newcommand {\chandra} {\textsl{Chandra}}

\newcommand {\nustar} {\textsl{NuSTAR}}
\newcommand {\suzaku} {\textsl{Suzaku}}
\newcommand {\integral} {\textsl{INTEGRAL}}

\def \rsun {\ifmmode$R$_{\odot}\else R$_{\odot}$}

          \font\sixrm=cmr6

\def\nudotW{\dot{\nu}_{\hbox{\sixrm W}}}
\def\nudotD{\dot{\nu}_{\hbox{\sixrm D}}}
\def\EdotD{\dot{E}_{\hbox{\sixrm D}}}

\def \hcm {\hbox {\ifmmode $ atoms cm$^{-2}\else atoms cm$^{-2}$\fi}}

\def\approxgt{\mathrel{\hbox{\rlap{\lower.55ex \hbox {$\sim$}}
        \kern-.3em \raise.4ex \hbox{$>$}}}}
\def\approxlt{\mathrel{\hbox{\rlap{\lower.55ex \hbox {$\sim$}}
        \kern-.3em \raise.4ex \hbox{$<$}}}}

\def \arcsec {\hbox{$^{\prime\prime}$}}

\def \src {SGR\,1806$-$20}

\begin{document}

\title{The Sleeping Monster: \nustar\ observations of \src, 11 years after the Giant Flare}

%
\author{George~Younes\altaffilmark{1,2}}
\author{Matthew~G.~Baring\altaffilmark{3}}
\author{Chryssa~Kouveliotou\altaffilmark{1,2}}
\author{Alice~Harding\altaffilmark{4}}
\author{Sophia~Donavan\altaffilmark{1,2}}
\author{Ersin~G\"o\u{g}\"u\c{s}\altaffilmark{5}}
\author{Victoria~Kaspi\altaffilmark{6}}
\author{Jonathan~Granot\altaffilmark{7}}

\altaffiltext{1}{Department of Physics, The George Washington University, Washington, DC 20052, USA, gyounes@gwu.edu}
\altaffiltext{2}{Astronomy, Physics and Statistics Institute of Sciences (APSIS), The George Washington University, Washington, DC 20052, USA}
\altaffiltext{3}{Department of Physics and Astronomy, Rice University, MS-108, P.O. Box 1892, Houston, TX 77251, USA}
\altaffiltext{4}{Astrophysics Science Division, NASA Goddard Space Flight Center, Greenbelt, MD 20771}
\altaffiltext{5}{Sabanc\i~University, Orhanl\i-Tuzla, \.Istanbul 34956, Turkey}
\altaffiltext{6}{Department of Physics, McGill University, Montreal, Quebec, H3A 2T8, Canada}
\altaffiltext{7}{Department of Natural Sciences, The Open University of Israel, 1 University Road, P.O. Box 808, Ra\'anana 43537, Israel}

\begin{abstract}

We report the analysis of 5 \nustar\ observations of \src\ spread over a year from April 2015 to April 2016, more than 11 years following its Giant Flare (GF) of 2004. The source spin frequency during the \nustar\ observations follows a linear trend with a frequency derivative $\dot{\nu}=(-1.25\pm0.03)\times10^{-12}$~Hz~s$^{-1}$, implying a surface dipole equatorial magnetic field $B\approx7.7\times10^{14}$~G.  Thus, \src\ has finally returned to its historical minimum torque level measured between 1993 and 1998. The source showed strong timing noise for at least 12 years starting in 2000, with $\dot{\nu}$ increasing one order of magnitude between 2005 and 2011, following its 2004 major bursting episode and GF. \src\ has not shown strong transient activity since 2009 and we do not find short bursts in the \nustar\ data. The pulse profile is complex with a pulsed fraction of $\sim8\%$ with no indication of energy dependence. The \nustar\ spectra are well fit with an absorbed blackbody, $kT=0.62\pm0.06$~keV, plus a power-law, $\Gamma=1.33\pm0.03$. We find no evidence for variability among the 5 observations, indicating that \src\ has reached a persistent and potentially its quiescent X-ray flux level after its 2004 major bursting episode. Extrapolating the \nustar\ model to lower energies, we find that the 0.5-10~keV flux decay follows an exponential form with a characteristic timescale $\tau=543\pm75$~days. Interestingly, the \nustar\ flux in this energy range is a factor of $\sim2$ weaker than the long-term average measured between 1993 and 2003, a behavior also exhibited in SGR~$1900+14$. We discuss our findings in the context of the magnetar model.

\end{abstract}

\section{Introduction}
\label{Intro}

Magnetars are a small class of isolated neutron stars believed to be
powered by the decay of their strong ($B\sim10^{14-16}$~G) internal
magnetic fields \citep[see][for reviews]{mereghetti08AARv:magentars,
  turolla15:mag,kaspi17:magnetars}. This characteristic induces very
peculiar observational properties to the class. Almost all magnetars
have been observed to enter bursting episodes where they emit 10s to
100s of short ($\sim0.1$~s), bright ($10^{37}-10^{41}$~erg), hard
X-ray bursts within the span of days to weeks \citep[e.g.,
][]{israel08ApJ:1900,lin11ApJ:1E1841,vanderhorst12ApJ:1550,gogus17ApJ}.
These episodes are usually accompanied by changes in the spectral and temporal
properties of the source persistent X-ray emission. The persistent X-ray
flux increases, occasionally by as many as three orders of magnitude
\citep[e.g.,][]{kargaltsev12apj:1834,scholz12ApJ:1822,rea11:outburst},
and their spectra harden. The shape of the pulse profile and the
fraction of the pulsed flux change, while the source timing properties
vary, either in the form of a glitch or a more gradual change in the
spin-down rate \citep[e.g.,][]{dib14ApJ,archibald15ApJ:1048}. The
source properties recover to pre-outburst levels months to years
later. Hence, observations of magnetar outbursts are a key ingredient
to understanding the physics behind these perplexing sources, and the
geometrical locale of their activity.

\src\ is historically the most active magnetar in the family, known to emit short
bursts regularly since its discovery. Major bursting episodes have been recorded several
times with the strongest one occurring from mid to late 2004. This episode
culminated with the emission of the strongest giant flare (GF) on
record so far \citep{hurley05NaturGF1806,gaensler05Natur:1806},
December 27th, 2004 (MJD 53366).
Radical changes in the source temporal and spectral properties have 
been observed since 1995 with its X-ray spectral shape hardening
gradually and its frequency derivative increasing monotonically up to
2002 \citep{mereghetti05ApJ:1806,woods07ApJ:1806}. Around the time of
the 2004 bursting episode and the giant flare, erratic changes to the
timing properties of the source were observed \citep{woods07ApJ:1806}.
The 0.5-10~keV persistent flux from the source started increasing shortly before
the major bursting episode of 2004, reaching a maximum around its peak
activity. The GF did not have a measurable effect on the spectral
properties of the source persistent emission, while it did decrease 
the pulsed fraction from its historical level of $8\%$ to about $3\%$
\citep{rea05ApJ:1806,tiengo05:1806}.

In \citet[][Y15 hereinafter]{younes15ApJ:1806}, we studied the X-ray
properties of \src\ up to mid 2011, over seven years following the
GF. We found that the torque on the star still showed strong variation
and, on average, remained at a historically high level, an order of
magnitude larger than the one measured between 1994 and 1998. The
pulse profile was double peaked with a modest contribution from a
second harmonic. The source flux started decreasing in 2005 towards its
quiescent value, while at the same time its blackbody (BB)
temperature $kT$ cooled and power-law (PL) slightly softened
(Y15). Here we report on the analysis of five \nustar\ observations of
\src\ spanning a full year from April 2015 to April 2016, over 11
years following its major bursting episode and GF. We present the
observations and data reduction in Section~\ref{obs}; the data
analysis and results are presented in Section~\ref{res}. Finally, 
both the temporal and spectra results are discussed in Section~\ref{discuss}
in the context of the magnetar paradigm, focusing on field structure 
and magnetospheric emission models.

\section{Observations and data reduction}
\label{obs}

The {\it Nuclear Spectroscopic Telescope Array} (\nustar, \citealt{
  harrison13ApJ:NuSTAR}) consists of two identical modules FPMA and
FPMB operating in the energy range 3-79~keV. \nustar\ observed \src\
on five occasions, the first of which took place on 2015 April 17. The
last observation was taken on 2016 April 12 (Table~\ref{logObs}). We
processed the data using the \nustar\ Data Analysis Software,
\texttt{nustardas} version v1.5.1. We analyzed the data using the
\texttt{nuproducts} task (which allows for spectral extraction and
generation of ancillary and response files) and HEASOFT version
6.19. We extracted source events around the source position using a
circular region with 45\arcsec radius, which maximized the S/N
ratio. Background events were extracted from an annulus around the
source position with inner and outer radii of 80\arcsec\ and
120\arcsec, respectively. Only in the first observation did we have
both modules strongly contaminated by stray light, whereas at most one
module showed stray light contamination in the following four
observations.

\begin{table}[t]
\caption{\nustar\ observations and source timing properties}
\label{logObs}
\newcommand\T{\rule{0pt}{2.6ex}}
\newcommand\B{\rule[-1.2ex]{0pt}{0pt}}
\begin{center}{
\resizebox{0.48\textwidth}{!}{
\begin{tabular}{l c c c c}
\hline
\hline
Observation ID   \T\B & Date & Exposure & $\nu$ (error) & PF (error) \\
                           \T\B &         &      (ks)     &    (Hz) &    (\%)   \\
\hline
30102038002$^a$&2015-04-17&33.2&\ldots&$<15$\\
30102038004&2015-06-29&28.7&0.129030(3)&9(2)\\
30102038006&2015-08-19&31.2&0.129023(2)&7(1)\\
30102038007&2015-11-11&45.7&0.129013(2)&8(2)\\
30102038009&2016-04-12&29.9&0.128994(4)&7(1)\\
\hline
\end{tabular}}}
\end{center}
\begin{list}{}{}
\item[{\bf Notes.}]$^{a}$ No pulse frequency measurement was possible
  due to strong contamination from stray light. Pulse fraction
  upper limit consistent with the rest of the measurements.
\end{list}
\end{table}

We also include in our analysis one \chandra\ observation taken on
2000 August 15, with a total exposure of 31~ks (obs. ID 746). The
source was placed on the ACIS S3 chip which is used in a 1/4 subarray
mode, reducing the read-out time to 0.8~s. The spectral analysis of
this observation was never reported in the literature, to our best of
knowledge, due to mild pile-up issues with a fraction of $\sim8\%$ of
total counts being piled-up \citep{kaplan02ApJ:1806}. To identify
the historical flux level from \src, here we perform spectral analysis of
this observation.  CIAO version 4.9 was employed for our data reduction
purposes. We extract source counts from a circle with radius of
2\arcsec\ centered on the best location from the source \citep{
  kaplan02ApJ:1806}. Background counts are extracted from an annulus
region with inner and outer radii of 10\arcsec\ and 20\arcsec,
respectively, centered on the same location as the source circular
region.  The ancillary and response files were created using the 
{\tt mkacisrmf} and {\tt mkarf} tools, respectively.  Two
methods were adopted to mitigate the pile-up problem in the observation. 
Since the pile-up fraction is relatively low, we added the \chandra\
\texttt{pileup} model (included in XSPEC, \citealt{davis01apjpile}) to
the full spectral model we use to fit the source spectrum
(Section~\ref{specana}). As a validation of this method, the source spectrum
was extracted from the pile-up free wings of the ACIS psf,
excluding the piled-up 1.2\arcsec\ central core. We find consistent
results between both methods. In Section~\ref{specana} we only report
the spectral results as derived using the full PSF while including the
\texttt{pileup} model in the fit.

The spectral analysis of the \nustar\ and \chandra\ data was 
performed using XSPEC version 12.9.0k \citep{arnaud96conf}. 
The photo-electric cross-sections of \citet{
  verner96ApJ:crossSect} and the abundances of \citet{wilms00ApJ} are
used throughout to account for absorption by neutral gas. We bin the
spectra to have a minimum of 5 counts per bin, and used the Cash
statistic (C-stat) in XSPEC for model parameter estimation and error
calculation. We used the \texttt{goodness} command for goodness of fit
estimation. We double checked our spectral analysis results by binning
the spectra to have a S/N of 7 (about 50 counts per bin) and using
the typical $\chi^2$ method. Both methods gave consistent results. For
all spectral fits, we added a multiplicative constant normalization
between FPMA and FPMB, frozen to 1 for the former and allowed to vary
for the latter to account for any calibration uncertainties between
the two instruments. We find that this uncertainty clusters around
$\sim5\%$. Finally, all quoted errors are at the $1\sigma$
level, unless otherwise noted.

\newpage

\section{Results}
\label{res}

\subsection{Timing analysis}
\label{timeana}

\begin{figure}[]
\begin{center}
\includegraphics[angle=0,width=0.48\textwidth]{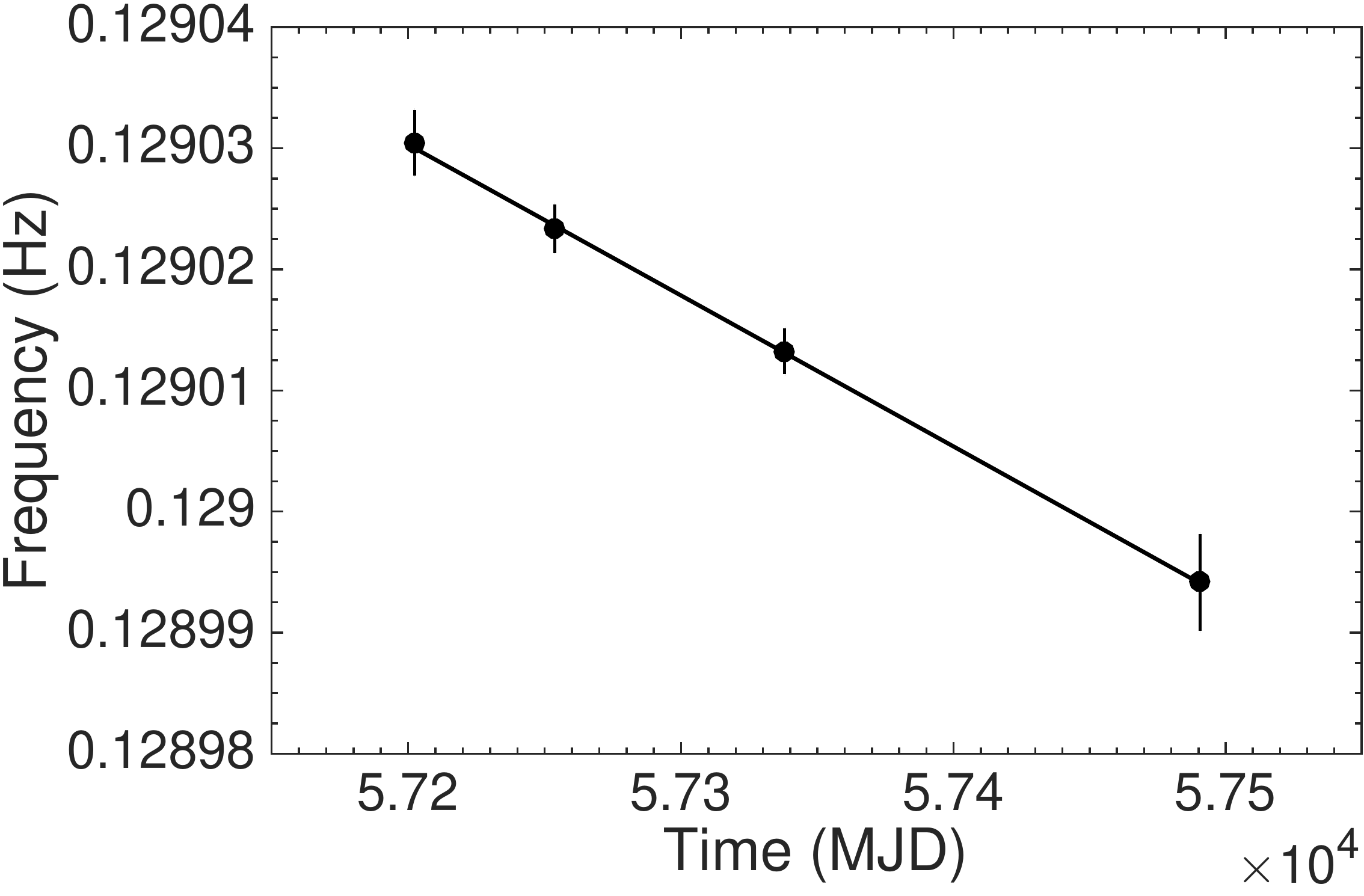}
\caption{Frequency (black dots) and best fit linear trend (solid line)
  to the four NuSTAR observations where the signal was detected. The
  slope of the best fit, hence frequency derivative, is
  $\dot{\nu}=(-1.25\pm0.03)\times10^{-12}$~Hz~s$^{-1}$.}
\label{freqHisNus}
\end{center}
\end{figure}

To maximize the S/N ratio for our timing analysis, we considered only
source events in the energy range 3-50 keV. We corrected these events 
arrival times to the solar barycenter and to drifts in the \nustar\
clock caused by temperature variations \citep{harrison13ApJ:NuSTAR}.
We applied the Z$^2_{\rm m=2}$ algorithm to search for the pulsations
from the source. We chose m=2 given the fact that the 2011 \xmm\
observations of the source still showed a double-peaked profile with
modest contribution from the second harmonic (Y15). We searched the
interval $0.126$-$0.130$~Hz with a size step of $2.0\times10^{-5}$~Hz,
which encapsulates the expected frequencies for the different
frequency derivatives that \src\ has shown since 1993. In all but the
first observation we detect a signal at around $8\sigma$ (trial
corrected). The results are given in Table~\ref{logObs} and shown in
Figure~\ref{freqHisNus}. The frequencies follow a linear trend
with a frequency derivative
$\dot{\nu}=(-1.25\pm0.03)\times10^{-12}$~Hz~s$^{-1}$. 

\begin{figure}[t]
\begin{center}
\includegraphics[angle=0,width=0.23\textwidth]{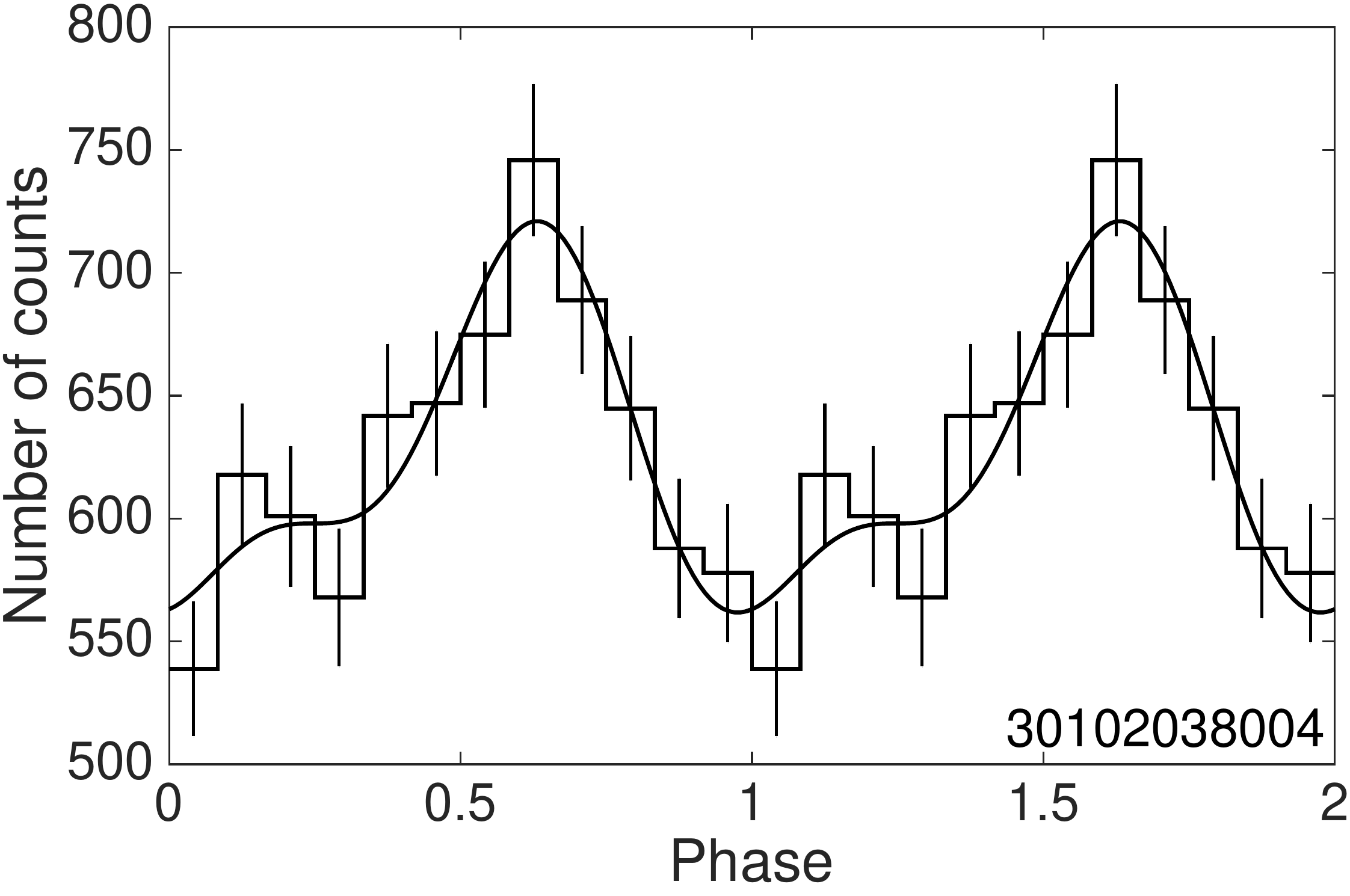}
\includegraphics[angle=0,width=0.23\textwidth]{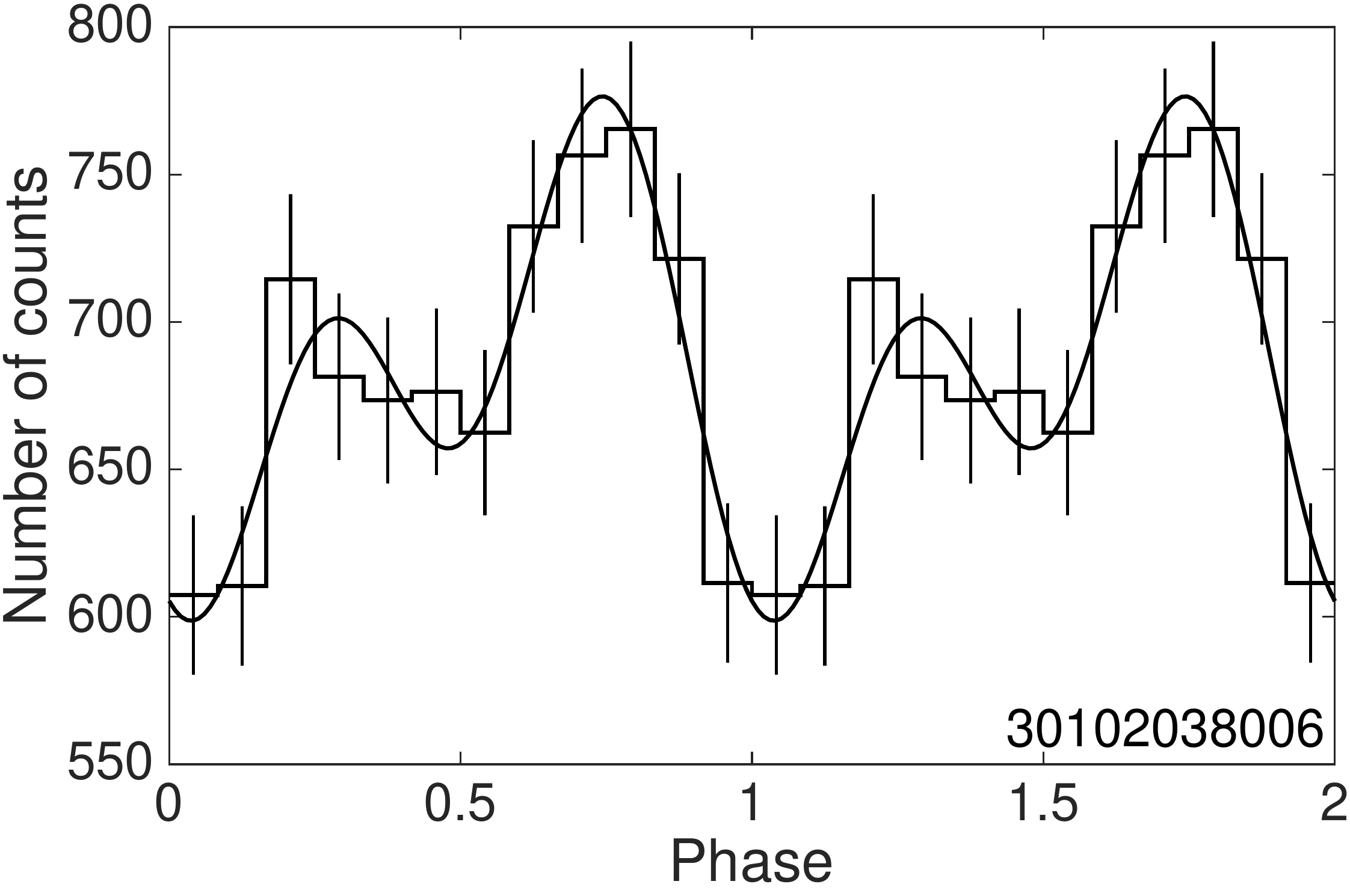}\\
\includegraphics[angle=0,width=0.23\textwidth]{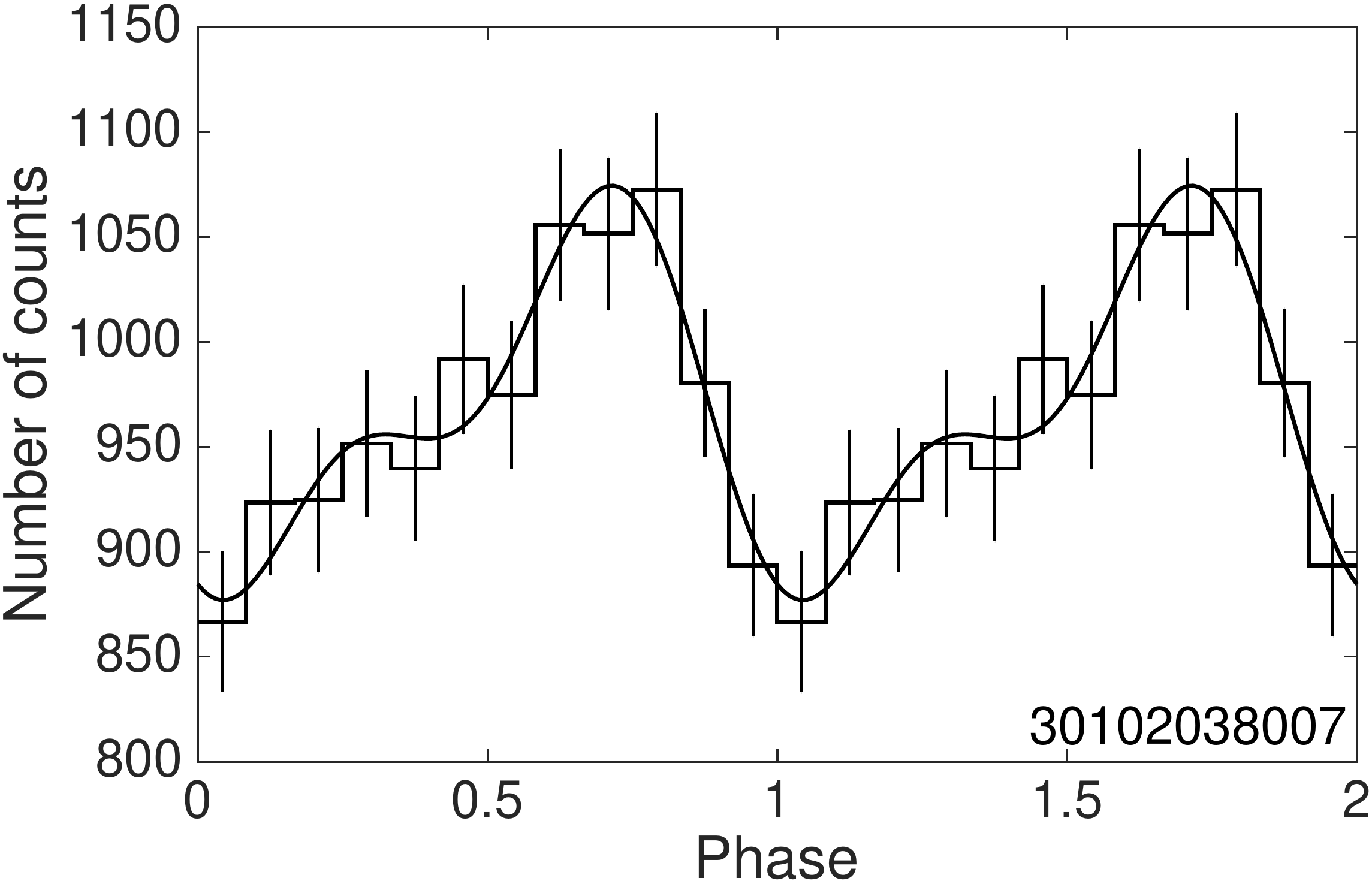}
\includegraphics[angle=0,width=0.23\textwidth]{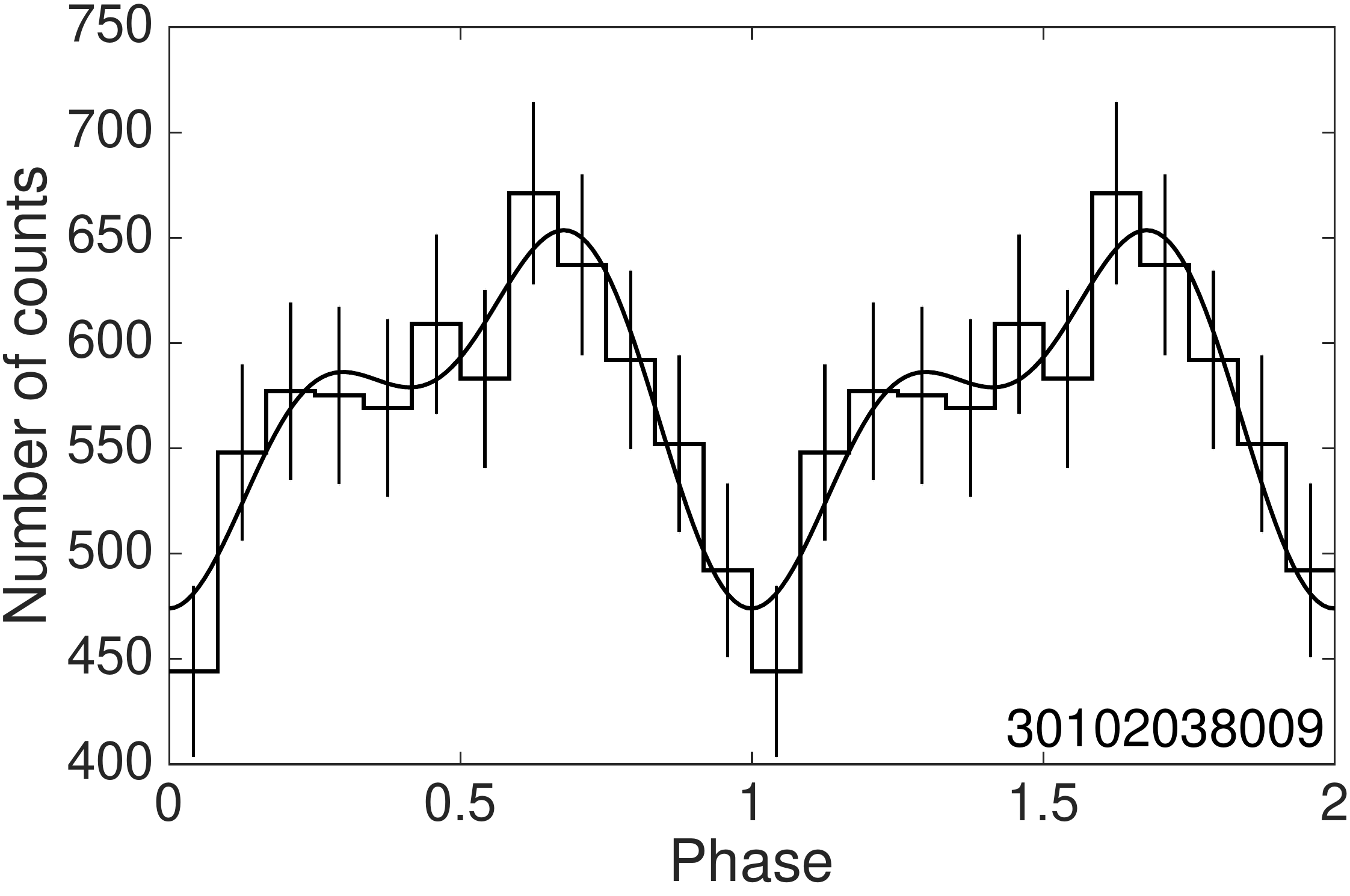}\\
\caption{The \nustar\ 3-50~keV background-corrected pulse profiles of
  \src\ where the pulse was detected, i.e., for the last four observing runs 
  listed in Table~\ref{logObs}. Two cycles are shown for
  clarity. The solid line is the best fit Fourier series including
  the contribution from 2 harmonics.}
\label{PPALL}
\end{center}
\end{figure}

We folded the event files of each of the last four observations
in the energy range 3-50~keV at their respective periods found
above, creating a pulse profile (PP), which we then
background-corrected (Figure~\ref{PPALL}). These PPs looked similar to
the ones following the 2004 GF, i.e., complex with a multi-peak
structure \citep[e.g., ][Y15]{mereghetti05ApJ:1806,woods07ApJ:1806}.
Therefore, we fit these PPs with a Fourier series including the
contribution from 2 harmonics \citep[e.g.,][Figure~\ref{PPALL}]{
  bildsten97ApJ:PP,younes15ApJ:1744}. The fits are good resulting in a
$\chi^2$ of $\sim5$ for 7 degrees of freedom (d.o.f.).

We estimated the rms pulsed fraction (PF) for these observations in
the energy range 3-50~keV. We find that the PF is stable at around
$8\%$. We also derive a 3$\sigma$ upper limit of $15\%$ on the pulsed
fraction of a fiducial signal for observation 1. These PFs level are
consistent with the historical level as measured with \asca\ and \sax\
around 1995. The only change in PF for \src\ was observed immediately
after the GF when it dropped to a minimum of 3\%\ \citep{rea05ApJ:1806,tiengo05:1806}.  
We searched for any changes in pulse morphology
and/or pulsed fraction with energy by considering events above and
below 10~keV separately. We find no dependence, within error, in these
two properties within the energy range considered.

\begin{figure*}[]
\begin{center}
\includegraphics[angle=0,width=0.88\textwidth]{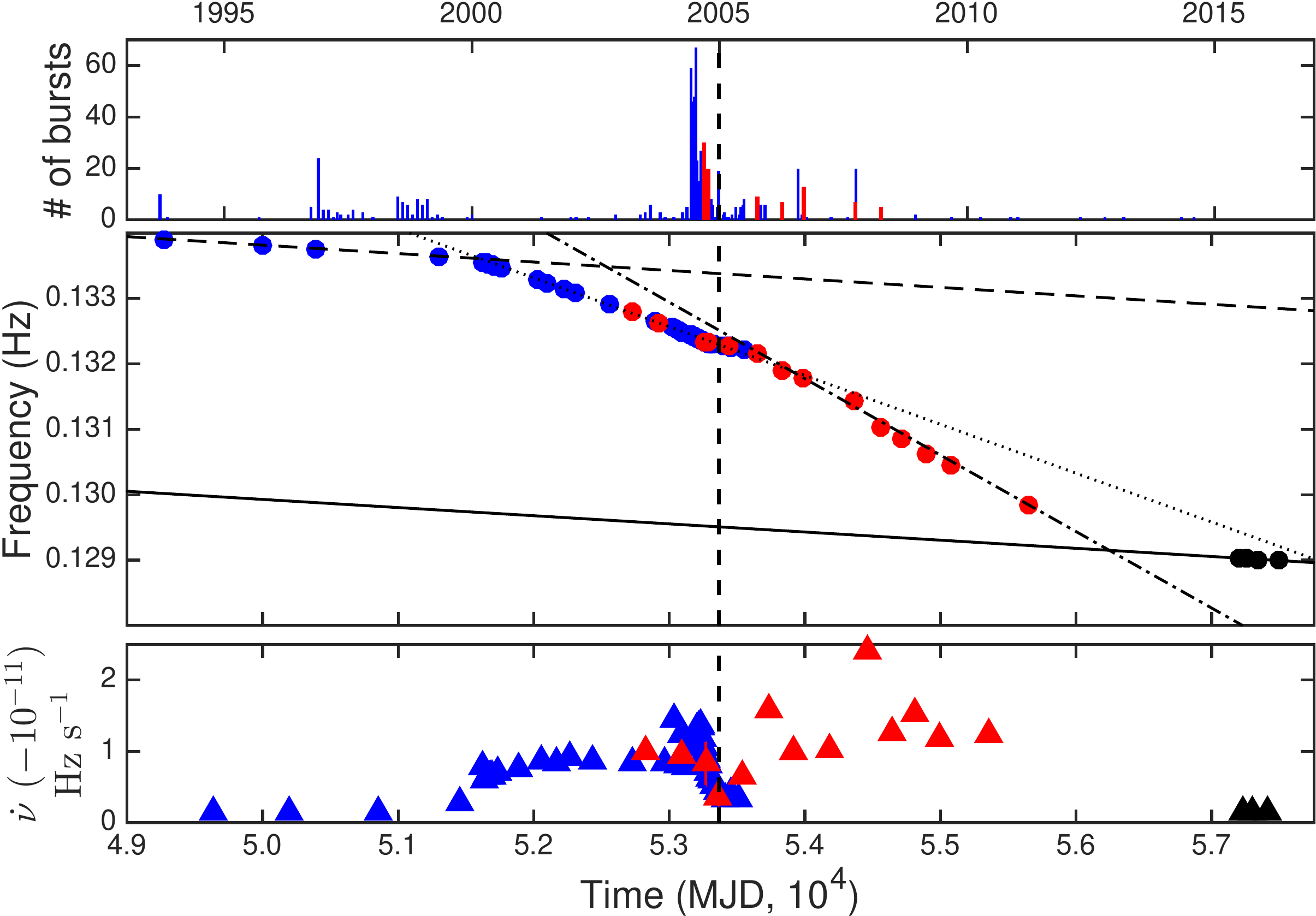}
\caption{Extension of the works by \citet{woods07ApJ:1806} and \citet[][Y15]{younes15ApJ:1806}
  showing the frequency and frequency derivative history of \src\ from mid
  1993 until April 2016. {\sl Top panel.} Number of bursts (per 30
  days). Data for the blue bars are collected from \citet{woods07ApJ:1806}, 
  Y15, including bursts reported in GCNs from 2012 to 2016. 
  Red bars represent bursts as detected by \xmm\ (Y15). The dashed
  vertical line marks December 27th, 2004, the date of the giant flare.
  {\sl Middle panel.} Spin frequency history. Blue dots are adopted
  from \citet{woods07ApJ:1806}, representing data from 
  five different X-ray telescopes, red dots are from Y15, while black
  dots are the \nustar\ frequency measurements. The vertical dashed
  line in all three panels denotes the time of the GF. The
  dashed and dotted lines are fits to the frequency derivative from
  1993 to 2000 January ($\dot{\nu}=-1.48\times10^{-12}$~Hz~s$^{ -1}$),
  and 2001 January to 2004 April
  ($\dot{\nu}=-8.69\times10^{-12}$~Hz~s$^{ -1}$,
  \citealt{woods07ApJ:1806}). The dot-dashed line is the fit to
  frequency measurements from 2005 July up to April 2011 
  ($\dot{\nu}=-1.35\times10^{-11}$~Hz~s$^{ -1}$, Y15). The black solid
  line is the best fit linear trend to the \nustar\ data
  ($\dot{\nu}=-1.25\times10^{-12}$~Hz~s$^{-1}$). {\sl 
    Bottom panel.} Instantaneous frequency derivative between two
  consecutive frequency measurements. Note the return of the
  instantaneous frequency derivatives at the time of the \nustar\
  observations to the 1995 historical level. See text for more
  details.}
\label{freqHis}
\end{center}
\end{figure*}

We extended the work of \citet{woods07ApJ:1806} and Y15 to build the
most comprehensive view of the torque evolution of \src\ from 1993 up
to 2016. The middle panel of Figure~\ref{freqHis} shows the timing
history of \src\ up to the last \nustar\ observation in April 2016,
over 11 years after the giant flare. The blue dots are data from
\citet{woods07ApJ:1806}, red dots are \xmm\ data from Y15, while black
dots are the frequencies as derived with \nustar. The lines represent
average frequency derivative over periods of relatively stable spin.
The bottom panel shows the instantaneous frequency derivative
calculated between two adjacent frequency data points (blue triangles
are data taken from \citealt{woods07ApJ:1806}, red triangles from Y15,
and black are for \nustar). Both the instantaneous and the average
frequency derivatives as derived with \nustar\ data show that the source has 
returned to a level consistent with its historical level,
e.g., $\dot{\nu}=(-1.22\pm0.17)\times10^{-12}$~Hz~s$^{-1}$ between
1996 November 5 and November 18 \citep{woods00ApJ:1806}. Assuming a similarly abrupt change in the frequency trend as seen with the other two changes,  the extrapolation of the \nustar frequency points (solid line) indicates that this change may have likely started around mid 2012. 

Similar to the previous works, we also report on the burst history
from the source from 2011 up to end of 2016 (Figure~\ref{freqHis}, top
panel). These are bursts reported in GCNs (Gamma-ray Coordinates
Network) and mainly seen with wide field-of-view instruments within
the InterPlanetary Network, IPN. It is evident that the source has
been in a quiet state with no major bursting episode since about
2009. Finally, we note that we searched all \nustar\ data for short
bursts using the method of \citet{gavriil04ApJ:1E2259} in the energy
range 3-79~keV. We used multiple time bins (16, 32, 64,
128, and 256 ms); we found no evidence of low level bursting activity in
\src. All the above results are discussed below in
Section~\ref{discuss}.

\subsection{Spectral analysis}
\label{specana}

We fit all \nustar\ spectra simultaneously with an absorbed PL and BB
model. We link the absorption between the 5 observations while we keep 
all other parameters free to vary. We find a good fit with a C-stat of
5505 for 5347 d.o.f. (\texttt{goodness} $\approx62\%$). We find an
absorption column density $N_{\rm
  H}=(1.0\pm0.3)\times10^{23}$~cm$^{-2}$, consistent with \xmm\  
spectral results (Y15). The rest of the parameters, i.e., the BB
temperature, the PL photon index, and their respective fluxes are 
consistent between all observations within errors
(Table~\ref{specParam}). Hence, to obtain a representative 
average between the observing runs, we refit all spectra simultaneously
linking all parameters between the spectra. We find an equally good
fit with  a C-stat of 5551 for 5369 degrees of freedom d.o.f
(\texttt{goodness} $\approx59\%$). This result indicates that the
persistent X-ray emission of source is currently in a steady state.

\begin{figure*}[t]
\begin{center}
\includegraphics[angle=0,width=0.49\textwidth]{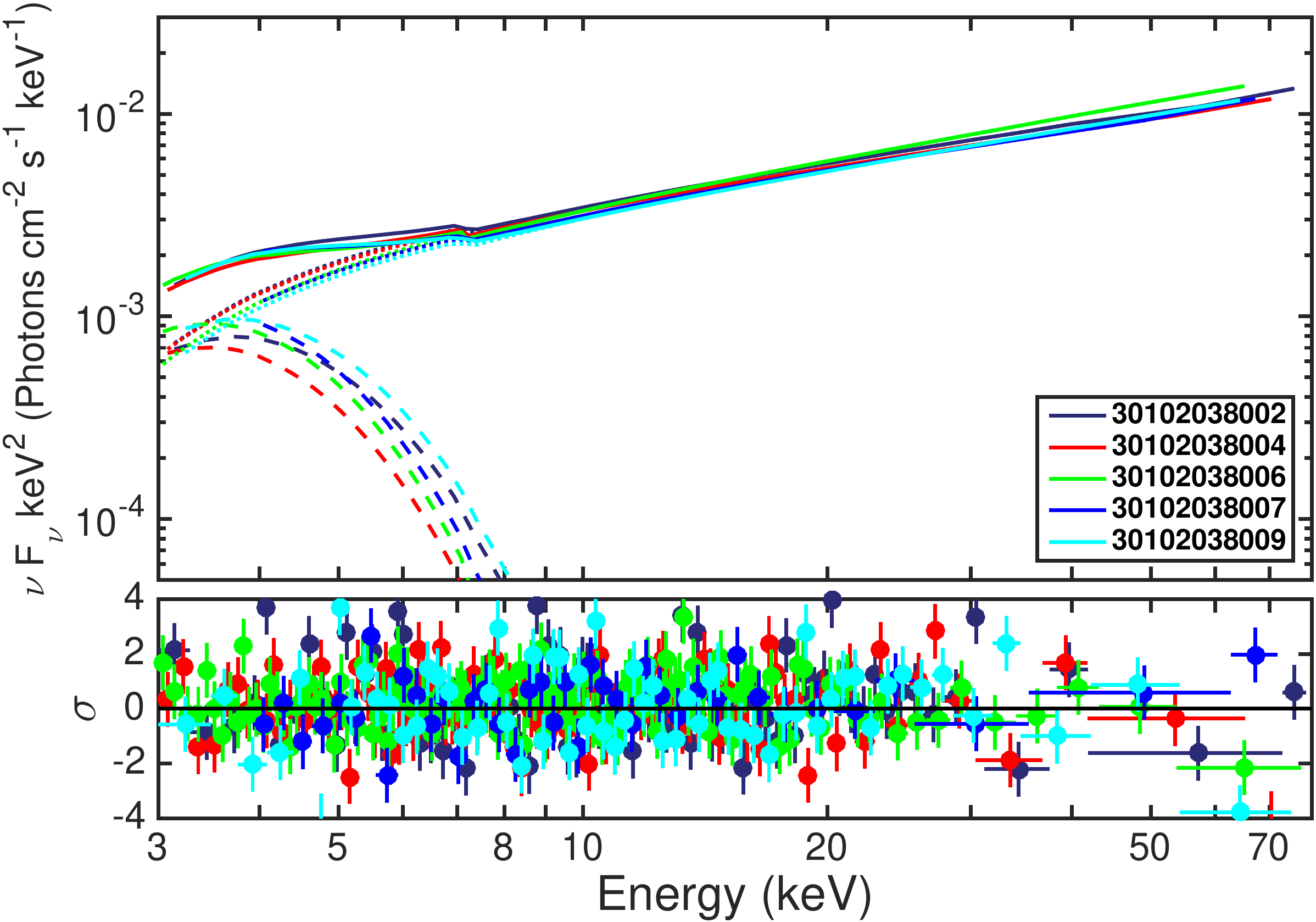}
\includegraphics[angle=0,width=0.49\textwidth]{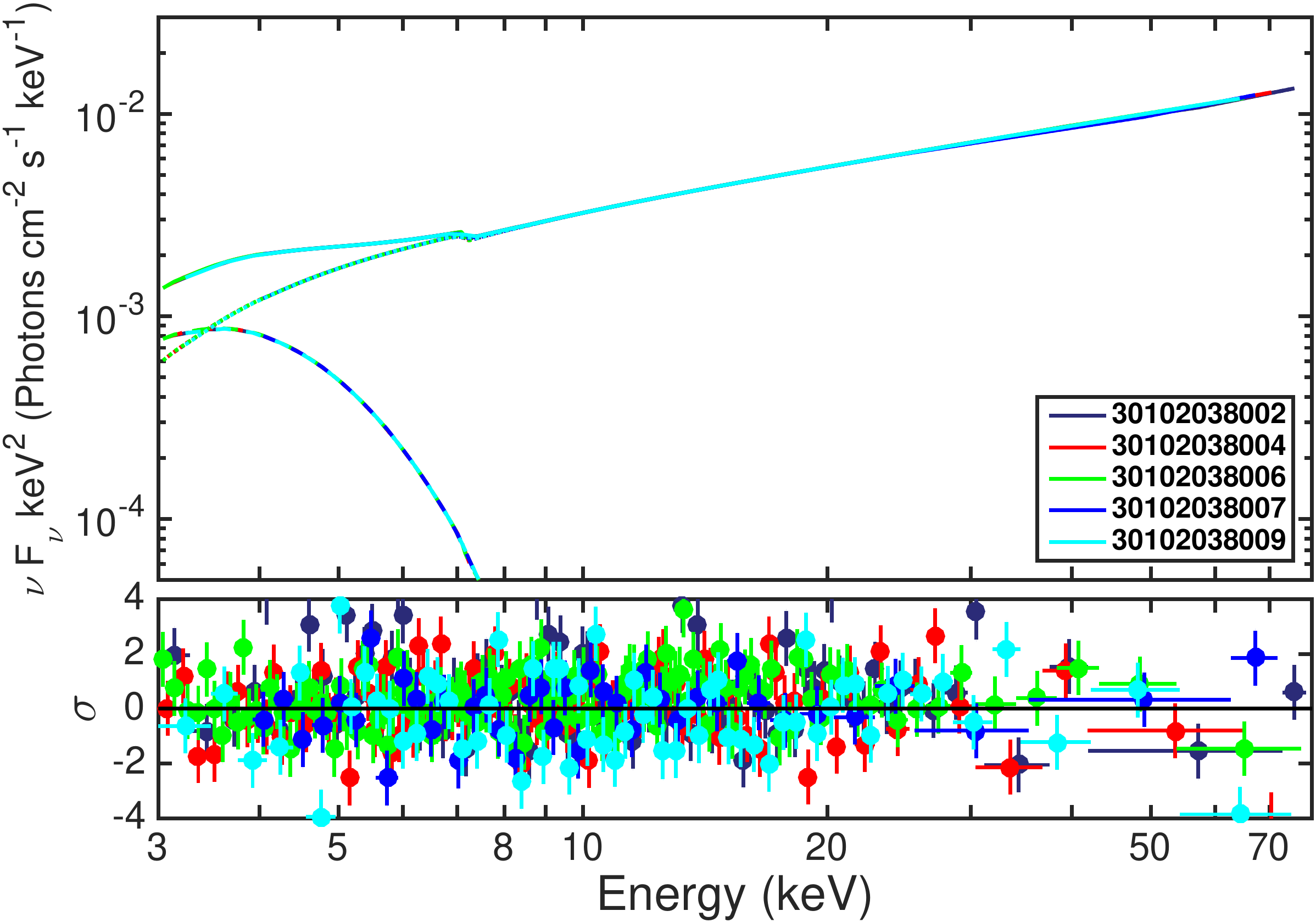}
\caption{BB+PL best fit model to all \nustar\ observations shown in
  $\nu F\nu$ space. {\sl Left Panels.} Parameters are left free to
  vary between all observations, except for the absorption column
  density. {\sl Right Panels.} All parameters linked between the 5
  observations to provide an ensemble average determination for them. 
  The {\sl upper panels} show the unfolded BB (dashed
  lines) and PL (solid lines) components. Data points were removed for
  clarity. The {\sl lower panels} show the residuals in terms of
  the standard deviation $\sigma$.}
\label{specFit}
\end{center}
\end{figure*}

We find a PL photon index $\Gamma=1.33\pm0.03$, a BB temperature
$kT=0.62\pm0.06$~keV, and, assuming a spherical surface for the
thermally emitting BB region, a radius $R=1.5\pm0.4$~km
(Table~\ref{specParam}). For an orthogonal rotator, 
a surface hot spot with this radius would clearly 
evince a higher pulse fraction than is presented in Fig.~\ref{PPALL}; 
more aligned magnetic and spin axes reduce the expected 
pulse fraction accordingly.  The 3-79~keV flux is
$(3.07\pm0.04)\times10^{-11}$~erg~s$^{-1}$~cm$^{-2}$. Extending the
\nustar\ model down to 0.5~keV we estimate a 0.5-79 keV flux of
$(3.68\pm0.05)\times10^{-11}$~erg~s$^{-1}$~cm$^{-2}$ and a luminosity
$(3.33\pm0.06)\times10^{35}$~erg~s$^{-1}$, assuming a distance of
8.7~kpc \citep{bibby08MNRAS:1806}. In this energy range, we find a BB
flux $F_{\rm BB}=(4.8\pm1.7)\times10^{-12}$~erg~s$^{-1}$~cm$^{-2}$ and
a PL flux $F_{\rm PL}=(3.21\pm0.04)\times10^{-11}$~erg~s$^{-1}$~cm$^{-2}$
implying a ratio $F_{\rm PL}/F_{\rm BB}\approx7$ (we note the weak 
constraint on the BB flux due to the lack of \nustar\ sensitivity at
energies below 3~keV).

\begin{table*}[t]
\caption{Spectral parameters for the BB+PL best fit model}
\label{specParam}
\newcommand\T{\rule{0pt}{2.6ex}}
\newcommand\B{\rule[-1.2ex]{0pt}{0pt}}
\begin{center}{
\resizebox{0.95\textwidth}{!}{
\begin{tabular}{c c c c c c c c}
\hline
\hline
Observation ID   \T\B & $N_{\rm H}$ & $kT$     & $R^{a}$       & $\log F_{\rm BB}$ & $\Gamma$ & $\log F_{\rm PL}$ & $\log F_{\rm tot}$ \\
                           \T\B & $10^{22}~$cm$^{-2}$ & (keV) & (km) & (erg s$^{-1}$ cm$^{-2}$) & & (erg s$^{-1}$ cm$^{-2}$)  & (erg s$^{-1}$ cm$^{-2}$)\\
\hline
\multicolumn{8}{c}{\nustar\ observations, parameters free to vary}\\
\hline
30102038002 \T\B & $10.0\pm3.0$ & $0.67_{-0.09}^{+0.10}$ & $1.1_{-0.4}^{+0.7}$ &$-11.63_{-0.18}^{+0.16}$ & $1.36\pm0.05$ & $-10.50\pm0.01$ & $-10.47\pm0.01$ \\
30102038004 \T\B & (L) & $0.59_{-0.07}^{+0.09}$ & $1.6_{-0.6}^{+1.0}$ & $-11.65_{-0.19}^{+0.18}$ &$1.38\pm0.04$ & $-10.53\pm0.01$ & $-10.50\pm0.01$ \\
30102038006 \T\B & (L) &  $0.59_{-0.06}^{+0.07}$ & $1.8_{-0.7}^{+0.9}$ & $-11.54_{-0.16}^{+0.15}$  &$1.27\pm0.04$ & $-10.48\pm0.01$ & $-10.45\pm0.01$ \\
30102038007 \T\B & (L) &  $0.61_{-0.06}^{+0.08}$ & $1.7_{-0.6}^{+1.0}$ & $-11.50\pm0.15$&$1.34_{-0.05}^{+0.04}$& $-10.53\pm0.01$ & $-10.49\pm0.01$ \\
30102038009 \T\B & (L) &  $0.67_{-0.09}^{+0.10}$ & $1.2_{-0.4}^{+0.8}$ &$-11.54_{-0.15}^{+0.14}$& $1.31\pm0.05$&$-10.54\pm0.01$ & $-10.50\pm0.01$ \\
\hline
\multicolumn{8}{c}{\nustar\ observations, parameters linked between all observations}\\
\hline
 \T\B & $10.0\pm2.0$  & $0.62\pm0.06$ & $1.5\pm0.4$ &$-11.57\pm0.15$& $1.33\pm0.03$&$-10.518\pm0.006$ & $-10.497\pm0.005$ \\

 \T\B &                          &                          &                       &$-11.32\pm0.15^b$&                         &$-11.14\pm0.04^b$ & $-10.92\pm0.04^b$ \\

\hline
\multicolumn{8}{c}{\chandra\ observation, 2000 August 15}\\
\hline
743 \T\B & $10.0\pm1.0$  & $0.6_{-0.1}^{+0.16}$ & $1.9_{-0.6}^{+1.1}$ &$-11.14_{-0.3}^{+0.2}$& $1.2_{-0.3}^{+0.5}$&$-10.82\pm0.09$ & $-10.64_{-0.08}^{+0.05}$ \\
\hline
\end{tabular}}}
\end{center}
\begin{list}{}{}
\item[{\bf Notes.}]$^{a}$Derived by adopting an 8.7~kpc distance
  \citep{bibby08MNRAS:1806}. \nustar\ fluxes are derived in the
  2-79~keV range, except for $^b$ which are derived in the energy
  range 0.5-10~keV. \chandra\ fluxes are derived in the 0.5-10~keV.
  Listed uncertainties are at the $1\sigma$ level.
\end{list}
\end{table*}

We fit the \chandra\ spectrum with the same model as the \nustar\ data
of a BB+PL (including the XSPEC \texttt{pileup} model, see
Section~\ref{obs}). We find a good fit with a C-stat of 498 for 512
degrees of freedom d.o.f. (\texttt{goodness} $\approx43\%$). The best
fit parameters are summarized in Table~\ref{specParam}. We find a
total 0.5-10~keV flux of $2.3_{-0.4}^{+0.3}\times10^{-11}$~erg~s$^{-1}$~cm$^{-2}$.

\begin{figure}[t!]
\begin{center}
\includegraphics[angle=0,width=0.48\textwidth]{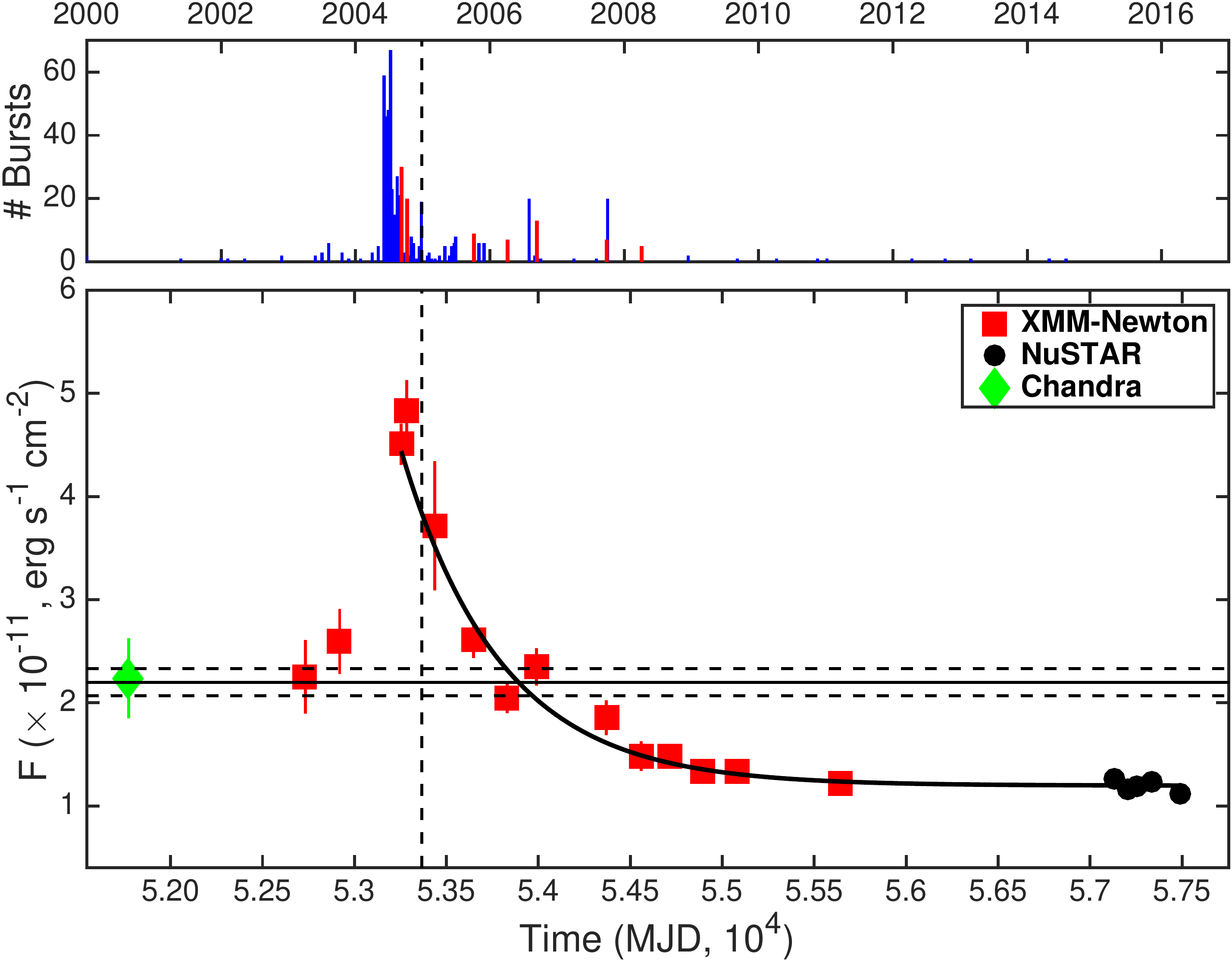}
\caption{Total, absorption-corrected, 0.5-10~keV flux evolution of
  \src\ since 2000. The solid line is an exponential decay fit to the
  data from the time of the first \xmm\ observation post-outburst, MJD
  53254 (2004 September 06), to the last \nustar\ observation, MJD
  57490 (2016 April 12). The characteristic decay timescale is
  $\tau=543\pm75$~days. Again, the dashed vertical line marks the date
  of the giant flare, MJD 53366 (December 27th, 2004).  The horizontal
  solid line marks the historical flux level between 1993 and 2001,
  with the parallel dashed lines defining the  $1\sigma$
  deviation. See text for details. Color coding for the bars and
  points is as in Figure~\ref{freqHis}.}
\label{fluxEvol}
\end{center}
\end{figure}

Figure~\ref{fluxEvol} shows the evolution of the total,
absorption-corrected, 0.5-10 keV flux of \src\ from 2000 up to the 
last \nustar\ observation. We also include the source average flux
measured from 1993 to 2001 with \asca\ and \sax\ (horizontal solid
line, \citealt{woods07ApJ:1806}). The enhancement of quiescent 
emission above its long-term persistent level reached its peak around the
\xmm\ observation of 2004 September 06 (observation ID 0205350101,
\citealt{woods07ApJ:1806}, Y15). We fit the flux evolution starting
at this data point with an exponential decay function of the form
$F(t)=K e^{-(t-t_{0})/\tau}+F_{\rm per}$. Here, $K$ is normalization,
$t_{0}$ is the time of the above \xmm\ observation, $\tau$ is the mean
lifetime for which the normalization decays by $63\%$, and $F_{\rm per}$ 
is the constant persistent flux level assumed to be the one
measured with \nustar. We find a good fit with a reduced $\chi^2$ of 1.2
for 15 d.o.f., with $\tau=543\pm75$~days and
$K=(3.2\pm0.4)\times10^{-11}$~erg~s$^{-1}$~cm$^{-2}$. The time for 
which the normalization decays by $95\%$, i.e., the flux has decayed
back to 5\% of its persistent level, is $1629\pm225$~days. The total
energy emitted in the outburst during this time interval is
$E=(1.4\pm0.4)\times10^{43}$~erg. 

The striking observational result in Figure~\ref{fluxEvol} is the
noticeable difference in the total 0.5-10~keV flux in the recent
\nustar\ flux compared with the pre-outburst long-term average, which
persisted from 1993 to 2003. The ratio of the 0.5-10~keV \nustar\ flux
to the historical average flux is $0.53\pm0.10$. The \nustar\ fluxes
derived for each model component as compared to those of \chandra\
(Table~\ref{specParam}) imply that the PL component has decreased
beyond its flux measured pre-outburst. On the other hand, the high
uncertainty on the BB flux measurement prevents us from drawing firm
conclusions on whether the BB component is behaving similarly to the
PL one. Nevertheless, combining our results with all the values
reported in the literature (\citealt{woods07ApJ:1806}; Y15) we find
that neither the BB temperature nor the PL index show any clear trend
in evolving from historical pre-outburst values to those determined
during the recent quiescent epoch.

\section{Summary and discussion}
\label{discuss}

\subsection{Timing Evolution}

Our \nustar\ observations spread over a year from April 2015 to April
2016 reveal the return of \src\ spin derivative,
$\dot{\nu}=(-1.25\pm0.03)\times10^{-12}$~Hz~s$^{-1}$, to its historical
minimum level derived more than 16 years earlier, e.g., 
$\dot{\nu}=(-1.22\pm0.17)\times10^{-12}$~Hz~s$^{-1}$ between 1996
November 5 and November 18 \citep{woods00ApJ:1806}. During the time
in between, \src\ showed radical changes in its temporal properties
while also being the most consistently burst-active magnetar.
It showed a major bursting episode in 2004 and several moderate ones
(10s of bursts), almost yearly from 1997 until 2009
(\citealt{woods02ApJ:1900}; \citealt{woods02ApJ:1900};
\citealt{mereghetti05ApJ:1806}; \citealt{woods07ApJ:1806};Y15). Since
2009, the source has been uncharacteristically quiet, with only
a few single bursts detected every year (Figure~\ref{freqHis}). Hence, this
level of torque that we derive with \nustar\ can be considered the
quiescent state magnetic configuration of \src. Assuming that this
corresponds to its dipole magnetic radiation, we estimate a magnetic
field strength at the equator $B=7.7\times10^{14}$~G, close in value
to those of SGR~1900+14 \citep{woods99ApJ:1900,mereghetti06ApJ:1900}
and 1E~1841$-$045 \citep{dib14ApJ}. We also estimate a spin-down age
$\tau=1.6$~kyrs, and note that because the torque evolution 
over the last 11 years has been so profound, it is clear that such spin-down 
ages are not an excellent proxy for the true stellar age. 
The last 2 \xmm\ observations indicate that the
source was still at a historically high $\dot{\nu}$ level in 2011. 
Due to the lack of observations between 2011 and 2015
we cannot exactly track the recovery of the spin down from the
source in transitioning from the high $\dot{\nu}$ state to 
the perennial one. However, if we
conjecture that the return to the minimum level is related to the lack
of bursting activity, with the last moderate bursting episode
occurring in 2009, we place a lower limit on the recovery timescale
of $\sim2$~yrs. This estimate agrees with the projected time of the torque change (mid 2012) mentioned in Section 3.1.

Similar to \src, XTE~J1810$-$197, the first so-called transient
magnetar \citep{ibrahim04ApJ:xtej1810}, is another source in the class
to have returned to a historical minimum level after displaying strong
timing anomalies following an outburst. A few months after the onset
of its 2003 outburst, $\dot{\nu}$ reached a factor of 8 larger than
its minimum observed value \citep{halpern05ApJ:xte1810,
  bernardini09:axp1810}. Interestingly, the source frequency
derivative returned back to its pre-outburst minimum around 2007, four
years after the outburst, and remained there until mid 2014 \citep{
  pintore16MNRAS:1810,camilo16ApJ:1810}. 1E~1048.1$-$5937 also shows
variation in $\dot{\nu}$ following its quasi-periodic outbursts,
sometime as large as a factor of 10. The torque then returns to its
nominal value on a timescale $\gtrsim1$~year \citep{
  archibald15ApJ:1048}. On the other hand, SGR~J1745$-$2900, the
Galactic center magnetar that went into outburst in April 2013
\citep{kennea13ApJ:1745,mori13ApJ:1745} has shown an increase of
$\dot{\nu}$ by a factor of 4.5, and no sign of decrease 3.5 years
following the outburst \citep{kaspi14ApJ:1745,zelati17MNRAS:1745}.
Last but not least, SGR~1900$+$14 has also shown strong timing noise
following burst-active episodes, e.g., with $\dot \nu$ increasing by a
factor of 5 following its late 1998 major bursting episode and GF
\citep{woods02ApJ:1900,mereghetti06ApJ:1900}. There is no published
information on the source $\dot \nu$ following its 2006 major bursting 
episode, hence, at the current time its $\dot \nu$ fate remains
unknown (Younes et al. in prep.). Due to the scarce data and the low
number of sources, it is not possible yet to do a systematic
comparison between the different objects. Nonetheless, such enhanced
spin-down post-outburst is common in magnetars, even those with
different levels of bursting behavior, and with different recovery
timescales.

The evolution of the timing signatures over durations spanning a few 
to ten years is relevant to the transient powering of magnetar
magnetospheres, both prior to and subsequent to bursting activity.
The leading model for activation of closed field regions in magnetars
considers dynamic, twisted magnetospheres that generate electric fields
and currents, a concept proposed by \citet{thompson02ApJ:magnetars}
for the quiescent emission, and embellished upon by \citet{
  beloborodov07ApJ:magCorona}.  Departures from dipolar field geometry
by small twist angles $\Delta \varphi\ll 1$ are invoked, and these
precipitate currents $\mathbf{j} \sim [c \boldsymbol{B}/(4 \pi r)]
\, \Delta \varphi \, \sin^2 \vartheta $ at magnetic colatitude
$\vartheta$ that generate electric field components $E_{\parallel}
\sim \sqrt{ 4\pi m_ec \,\vert \mathbf{j}\vert / e} $ parallel to the
local field \citep[see][for details]{beloborodov07ApJ:magCorona}.  The
ensuing acceleration can easily generate a hot corona that persists
for long activation times. In that paper, the resistive decay timescale
for the twist via ohmic dissipation couples both to the electric potential,
which is universally near the 1 GeV level, and also the X-ray
luminosity.  Thus, \citet{beloborodov07ApJ:magCorona} conclude that
twist dissipation activity in magnetars triggered by bursting activity
should last for timescales in the realm of several months to a few
years.

While this estimate is fairly close to the $e$-folding time for relaxation
that is inferred here from our timing results, the precise twist decay
timescale determination requires detailed simulational modeling.  The
recent developments of \citet{parfrey13ApJ} and \citet{chen17:mag}
forge steps in this direction, and in particular, the particle-in-cell
plasma simulations of \citet{chen17:mag} confirm that untwisting of
the magnetosphere does arise on ohmic dissipation timescales. Yet, in
this theory, the twists define a field morphology perturbation in the
inner magnetosphere, and it remains to be determined how and if such
structural deformations can account for the large torque changes that
must accompany the amplifications of ${\dot \nu}$ by a factor of 10
overall.

As an alternative origin, observe that
enhanced plasma loading of magnetar winds may contribute significantly
to the torque evolution.  The increase in $\dot \nu$ following periods
of bursting and the gradual return to a quiescent $\dot \nu$ and flux
level in SGR 1806-20 is consistent with the picture outlined by
\citet[][see also \citealt{tong13ApJ:wind}]{harding99ApJ:mag} of
magnetar bursts leading to episodic particle wind outflow that
temporarily increases the spindown rate, on top of a persistent
magnetic dipole spin evolution. From Eq.~(12) of
\citet{harding99ApJ:mag}, assuming that the wind luminosity $L_p$ is
much larger than the dipole spin-down power, $\EdotD$, then $L_p
\sim (\nudotW/\nudotD)^2 \EdotD$, where $\nudotW$ is the
enhanced frequency derivative following bursting periods, and 
$\nudotD$ is the frequency derivative of magnetic dipole spin down.
Adopting the historical frequency derivative $\nudotD = -1.22
\times 10^{-12}\,\rm Hz\,s^{-1}$ that is very close to the present
\nustar\ result (see Figure~\ref{freqHis} caption) to represent the
long-term value for the magnetic dipole torque, and using the
increased frequency derivative measured over the two periods, 
$\nudotW = -8.69 \times 10^{-12}\,\rm Hz\,s^{-1}$ (for 2001--2004) and
$\nudotW = -1.35 \times 10^{-11}\,\rm Hz\,s^{-1}$ (2005--2011), the
estimated wind luminosities producing the enhanced torque are $L_p
\sim 50.7\,\EdotD = 1.6 \times 10^{35}\,\rm erg\,s^{-1}$ (for
2001--2004) and $L_p \sim 122\,\EdotD = 3.8 \times 10^{35}\,\rm
erg\,s^{-1}$ (2005--2011) for the respective epochs.  

These $L_p$ values are
similar to the X-ray luminosity estimated in Section 3.2 from the spectral
fits of the NuSTAR data, indicating that the quiescent luminosity and
the enhanced particle wind power implied by the torque changes are
both around $100 \EdotD$.  This comparability may be 
coincidental, though a connection between wind power lost to infinity
and luminosity in trapped plasma that is dissipated in radiative form 
is naturally expected: the detailed nature of this coupling is 
not yet understood.  The increased particle flux following
bursting activity can deposit a large amount of energy in the
magnetar's environs. This possibility for transient powering of the
newly discovered nebula around magnetar Swift J1834.9-0846
\citep{younes12ApJ:1834,younes16ApJ:1834} was explored by
\citet{granot17:1834}.  It is therefore of significant interest how 
much particle power active burst episodes associated with GFs contribute 
to the cumulative, long-term energetics of a surrounding nebula.
In particular how such transient contributions compare with 
those of less dynamic and more prolonged strong wind 
epochs coupled with somewhat enhanced $\dot{\nu}$ values.

\subsection{Flux History}

The flux from the source has now reached a persistent level of the
order of $1.2\times10^{-11}$~erg~s$^{-1}$~cm$^{-2}$ in the 0.5-10~keV
range (after extrapolating the \nustar\ model to the lower end). The
flux decay prior to our observations follows a simple exponential
function with a characteristic timescale $\tau=543$~days. Such long
decay timescales have been seen in other magnetars
\citep[e.g.,][]{scholz14ApJ:1822,zelati17MNRAS:1745,alford16ApJ:1810}. We
refer the reader to Y15 for a detailed discussion of the
consequences of such long time recoveries. However, we will reiterate
here, that while both the spectral and temporal properties of \src\
have now reached a quiescent state, it is clear that they did not
follow the same long-term relaxation trend. The source X-ray flux
started decaying immediately following the peak of the 2004 outburst
(Y15, Figure~\ref{fluxEvol}), while the temporal properties lingered
at a large historical level between 2005 and 2011 when the source was
still moderately bursting. It reached a historically low level in
2015, following 6 years of burst-quiet period. This reinforces our Y15
conclusion that low level seismic activity causing small twists in the 
open field lines might be driving torque variations without having any
noticeable effects on the spectral behavior from the source.

The conspicuous result in the 0.5-10~keV flux recovery of \src\ after
the 2004 outburst is the lower quiescent level derived with \nustar\
compared with the long-term average pre-outburst, which persisted from
1993 to 2003 (Figure~\ref{fluxEvol}). The ratio of the 0.5-10~keV
\nustar\ flux to the historical average flux is $0.53\pm0.10$, i.e., a
factor of $\sim2$ smaller. A similar behavior was noticed in the flux
evolution of SGR~$1900+14$: the source flux prior to the 1998 GF as
measured over a 2-year period was at the
$1\times10^{-11}$\,erg~s$^{-1}$~cm$^{-2}$ level \citep{
  woods01ApJ:1900}, while the flux in 2005, after almost 3 years
during which no bursts were detected, reached half that value, i.e.,
$0.5\times10^{-11}$~erg~s$^{-1}$ cm$^{-2}$ \citep{
  mereghetti06ApJ:1900}, another example of a factor of 2 change in
the recovery to the apparent quiescent state.

These lower asymptotic fluxes relative to their respective historical
level constitute an interesting result. It is possible that this might
potentially be due to a reconfiguration of the internal magnetic field
in association with the lead-up to the GF. Changes in crustal field
morphology could affect the heat conduction between the hot neutron
star core and the surface; such a conductivity is extremely efficient
in polar zones where the magnetic field lines are oriented
approximately vertically. One might then expect heating of the surface
and also energy deposited in the magnetosphere approximately
contemporaneous with adjustments to field structure.  This might
explain the rising quiescent BB+PL fluxes during the main bursting
episode prior to the GF. The subsequent flux decline would signal an
ensuing cooling phase. A possible signature of a permanent
reconfiguration could be an alteration of the effective area of the BB
component.  Unfortunately, the uncertainty in the BB flux
determination from \nustar\ spectroscopy precludes clear inferences of
this possibility (see Table~\ref{specParam}), though there is a slight
hint of a net area reduction over the 15 year period.

\subsection{Spectral Models}

The non-thermal spectra obtained in our \nustar\ observations of \src,
embodied in Fig.~\ref{specFit}, are quite similar to the hard X-ray
tail components in other magnetars \citep[e.g.,
see][]{gotz06AA:1900,enoto10ApJ,vogel14ApJ:2259,tendulkar15ApJ:0142,
  younes17:1935,enoto17ApJS}. Yet, the power-law fit index of $\Gamma
= 1.33\pm 0.03$ we obtain is slightly flatter than the typical values
obtained in other observations of this source. \citet{
  mereghetti05ApJ:1806} derived $\Gamma = 1.5 \pm 0.3$ from 2004
\integral-IBIS observations just prior to the GF in December
that year. \citep{esposito07AA:1806} obtained $\Gamma = 2.0 \pm 0.2$
from \suzaku\ HXD-PIN data from September 2006 in the 10-40 keV range,
while \citep{enoto10ApJ} determined $\Gamma = 1.7 \pm 0.1$ with 2007
data from \suzaku.  A more recent summary of \suzaku\ observations for
SGR 1806-20 and other magnetars is presented in \citep{enoto17ApJS}.
Thus, while there was at first a suggestion of spectral flattening
associated with the lead up to the GF, the fits obtained here
indicate that there appears to be no clear evolutionary trend of the
power-law index during the recovery phase following that extreme
outburst.

The most popular paradigm for the generation of the hard X-ray
components in magnetars is resonant inverse Compton scattering
\citep{baring07,fernandez07ApJ}. Relativistic electrons, accelerated
in the inner magnetosphere in closed field line regions with highly
super-Goldreich Julian densities, up-scatter the abundant soft X-ray
photons emanating from magnetar atmospheres. This process is extremely
efficient since the scattering is resonant at the cyclotron energy
\citep[e.g.,][]{herold79:qed}, enhanced by over two orders of
magnitude relative to the Thomson value.  This can thereby effectively
convert electron kinetic energy into radiative form
\citep{baring11ApJ}. The kinematics of this process make for the
generation of flat spectra if the electrons are mono-energetic
\citep{baring07}, with quasi-power-law indices of $\Gamma \sim 0$ that
are of lower value in general than those for the typical hard X-ray observations. 
In the magnetic Thomson construction of \citet{beloborodov13ApJ},
integrating over emission volumes and limiting the maximum Lorentz
factor $\gamma_e$ of the electrons can generate emission spectra that
approximate magnetar hard X-ray components quite well, as is
demonstrated by the detailed comparison of models with spectral data
for three magnetars in \cite[][see also,
\citealt{an13ApJ:1841,vogel14ApJ:2259,an15ApJ:1841}]{hascoet14ApJ}.

Yet there is great complexity in full QED analyses of resonant Compton
upscattering spectra, as expounded in Wadiasingh et al. (2017) for
mono-energetic electrons. Therein, flat spectra from resonant scatterings
involving electrons moving along individual field lines, are steepened
when integrating over toroidal surfaces. Moreover, there is the
expectation that integrations over volumes within about 10 stellar
radii of the surface, and the introduction of electron cooling, will
soften these further to be approximately commensurate with the
$\Gamma$ values presented for \src\ here. Depending on the
observer viewing perspective, and the electron Lorentz factor, the
quasi-power laws can extend down into the soft X-rays below 3 keV.
Generally, this contribution is obscured by the thermal atmosphere component.
However, \src\ presents a special case in that the power-law tail
component blends closely into the thermal (BB) portion of the
spectrum, as is evident in Fig.~\ref{specFit}, which closely resembles
the BB+PL combination in Fig.~1 of \citet{enoto10ApJ}. This property
of an unusually high luminosity for the PL component (more so than for
other magnetars; see \citealt{gotz06AA:1900,enoto10ApJ}) provides a
significant constraint on resonant upscattering models that is yet to
be fully explored. It affords the prospect of probes of the emission
geometry and the values of the relativistic electron Lorentz factors
and number density.

So too does the pulse profile information in Fig.~\ref{PPALL}, which in
one particular epoch evinces a double-peaked profile. Wadiasingh et al.
(2017) illustrate how such double peaks can arise when the
viewing angle $\zeta$ and the magnetic axis angle $\alpha$ to the
rotation axis $\boldsymbol{\Omega}$ are similar in value, specifically
for emission from toroidal field surfaces.  In such cases, the line of
sight can sweep across quasi-polar regions as the star rotates.  This
temporal feature diminishes when the emission volume expands to
encapsulate a range of field line maximum altitudes and resonant
interaction locales, and the phase separation of the two peaks (see
Fig.~12 of Wadiasingh et al. 2017) declines with increasing photon
energy. Thus, as is the circumstance for gamma-ray pulsars, such pulse
profiles provide an important probe of the magnetic inclination
$\alpha$ of a magnetar, a prospect that is addressed in the \nustar\
analysis of data between 20 and 35 keV from 1E 1841-045
\citep{an13ApJ:1841,an15ApJ:1841}. More model development is needed to interpret
these properties with greater precision, and the observations we
present here set the scene to motivate such theoretical analyses.  We
anticipate that our results for \src\ here can help inform the
understanding of magnetar emission geometry, and the activation (and
its evolution) of the magnetosphere in the decades subsequent to GF events.

\section*{Acknowledgments}

This work made use of data from the \nustar\ mission, a project led by
the California Institute of Technology, managed by the Jet Propulsion
Laboratory, and funded by the National Aeronautics and Space 
Administration. We thank the \nustar\ Operations, Software and
Calibration teams for support with the execution and analysis of these
observations. This research has made use of the \nustar\ Data Analysis
Software (NuSTARDAS) jointly developed by the ASI Science Data Center
(ASDC, Italy) and the California Institute of Technology (USA). GY
acknowledges support from NASA under \nustar\ Guest Observer cycle-1
program 01166, proposal number 14-NUSTAR14-0030. M.G.B. acknowledges
the generous support of the NASA Astrophysics Theory Program through
grant NNX13AQ82. VMK receives support from an NSERC Discovery Grant,
an Accelerator Supplement and from the Gerhard Herzberg Award, an
R. Howard Webster Foundation Fellowship from the Canadian Institute
for Advanced Research, the Canada Research Chairs Program, and the
Lorne Trottier Chair in Astrophysics and Cosmology. JG acknowledges
support from the Israeli Science Foundation under Grant No. 719/14.

\end{document}